\documentclass{elsart}
\usepackage{natbib}
\usepackage{epsfig}

\def\etal{et~al.}
\def\spose#1{\hbox to 0pt{#1\hss}}
\def\lta{\mathrel{\spose{\lower 3pt\hbox{$\mathchar"218$}}
     \raise 2.0pt\hbox{$\mathchar"13C$}}}
\def\gta{\mathrel{\spose{\lower 3pt\hbox{$\mathchar"218$}}
     \raise 2.0pt\hbox{$\mathchar"13E$}}}

\begin{document}
\runauthor{P.N.Best}

\begin{frontmatter}
\title{Clustering around radio galaxies at $\mathbf{z \sim 1.5}$}
\author{Philip N. Best}

\address{IfA, Edinburgh}

\begin{abstract}
The importance of studying old elliptical galaxies at redshift $z \sim 1.5$ is
reviewed, considering both what can be learned by extending studies of the
evolution of cluster galaxy scaling relations to earlier cosmic epochs, and
the age--dating of old elliptical galaxies at high redshifts. Following this,
the first results are provided of an on--going project to find such distant
elliptical galaxies, through an investigation of the cluster environments of
powerful radio sources with redshifts $1.44 < z < 1.7$. These studies show a
considerable excess of red galaxies in the radio sources fields, with the
magnitudes ($K \gta 17.5$) and colours ($R-K > 4$) expected of old passively
evolving galaxies at the radio source redshift. The red galaxy overdensities
are found on two different scales around the radio sources; a pronounced
small--scale peak at radial distances of $\lta 150$\,kpc, and a weaker
large--scale excess extending out to 1 - 1.5\,Mpc. The presence and richness
of these red galaxy excesses varies considerably from source--to--source. An
interpretation of these results is provided.
\end{abstract}

\begin{keyword}
galaxies: clusters : general --- radio: galaxies --- galaxies:
evolution 
\end{keyword}
\vspace*{-0.4cm}
\end{frontmatter}
\vspace*{-0.4cm}

\section{Introduction}
One of the key questions facing astrophysics today concerns the epoch and mode
of formation of the most massive elliptical galaxies. Whilst the majority of
galaxies can be easily accommodated by a slow assembly of galaxies from
bottom-up, within the standard predictions of semi--analytical models of a
cold dark matter Universe \cite{kau98,col00}, some lines of evidence suggest
that an earlier and more dramatic formation mechanism may be necessary to
accommodate (some of) the most massive galaxies. In particular:\\
\noindent {\it 1. Powerful AGN at high redshifts:} the host galaxies of
powerful radio sources are the most massive galaxies known at early cosmic
epochs [e.g. 3], with stellar masses of a few times $10^{11} M_{\odot}$ at
redshifts $1 \lta z \lta 2$ \cite{bes98}. At higher redshifts, even if fueled
at the Eddington limit, central supermassive black holes of $10^{8-9}
M_{\odot}$ are required to produce the observed radio luminosity, suggesting
that massive galaxies are already in place at redshifts $z \gta 5$
\cite{bre99}.\\
\noindent {\it 2. The nature of sub-millimetre selected galaxies:} many of the
powerful sub-mm selected galaxies seen with SCUBA are objects at $z > 2$,
which are currently forming stars at hundreds or thousands of solar masses per
year [e.g. 6]\nocite{sma02}. The K$-z$ relation for these SCUBA galaxies
indicates that these are amongst the most massive objects at their epoch
\cite{dun02}. SCUBA galaxies therefore appear to be massive galaxies which are
forming the bulk of their stars in a single dramatic starburst at high
redshift.\\
\noindent {\it 3. Cluster galaxy scaling relations:} the very low scatter
found around the colour magnitude relation [e.g. 8]\nocite{ter02} and the
fundamental plane (a tight correlation between the surface brightness ($I_{\rm
e}$), characteristic radius ($r_{\rm e}$), and velocity dispersion ($\sigma$)
of cluster ellipticals, which can be roughly approximated as: $r_{\rm e}
\propto \sigma^{1.2} I_{\rm e}^{-0.8}$; e.g. 9)\nocite{jor96} of elliptical
galaxies in low redshift clusters imply a fairly high formation redshift of
these ellipticals, with little bluening from later starbursting activity. The
fundamental plane essentially measures the mass--to--light ($M/L$) ratio of
the galaxies (from $M \propto \sigma^2 r_{\rm e}$ and $L \propto I_{\rm e}
r_{\rm e}^2$ it can be written as $M/ L \propto M^{\approx 0.25}$); by
comparing clusters at different redshifts [e.g. 10,11],\nocite{kel97,dok01}
the mass--to--light ratio is found to evolve very slowly with redshift out to
$z \sim 0.8$, as $\Delta Log(M/L) \propto 0.4z$.  Further, the scatter around
neither the fundamental plane nor the colour--magnitude relation shows any
significant increase out to redshifts $z \sim 1$ \cite{dok01,sta98}. These
results both further support the argument that the elliptical galaxies must
have formed at a very early cosmic epoch, with limited later star formation
(although see \cite{dok01} for discussion of a possible `progenitor--bias'
selection effect).\\
\noindent {\it 4. The ages of massive ellipticals at high redshift:} Dunlop
et~al.  \cite{dun96} age--dated the old red elliptical galaxy 53W091 at
$z=1.55$ through a detailed investigation of its optical spectral energy
distribution. They derived an age of 3.5\,Gyr, comparable to the age of the
Universe at that redshift, again implying a very early formation
epoch. Although there has since been considerable debate as to the age of this
galaxy, and of the similar 53W069, Nolan et~al \cite{nol01} convincingly
defend this initial age estimate. Similar results are beginning to be obtained
for Extremely Red Objects (EROs; e.g \cite{cim02} and refs therein).

The identification of a sample of cluster elliptical galaxies at redshifts $z
\sim 1.5$ would provide a key opportunity to advance our understanding of the
epoch and mode of formation of these objects. This is an early enough epoch to
test the `progenitor--bias' model (which would allow later star formation in
cluster ellipticals), and would also offer a unique opportunity to compare,
through age--dating techniques, the ages of the most luminous elliptical
galaxies in the cluster with those of less luminous cluster ellipticals, and
therefore directly observe whether there is an age--luminosity relation for
cluster ellipticals. Redshift 1.5 is a critical cosmic epoch for these
studies: it is still sufficiently nearby that cluster ellipticals can still be
identified on the basis of multi--colour imaging in a practical observing
time, and is also an epoch where elliptical galaxies are still expected to be
plentiful. In this article, the first steps to providing such a sample of
elliptical galaxies are discussed.

\section{Multi--colour imaging of AGN fields}
\vspace*{-0.1cm}

Beyond redshift $z \sim 1$, the two well--known methods for identifying galaxy
clusters, deep optical\,/\,near--IR imaging for galaxy overdensities, and
X--ray detections of the hot intracluster medium, both start to become very
expensive in terms of telescope time to achieve sufficient sensitivity for a
wide survey area. For this reason, an alternative method has frequently been
adopted: using targetted studies towards powerful AGN. At low redshifts AGN
are often found in relatively rich environments [e.g. 16]\nocite{hil91}, and
at redshifts $z \sim 1$ numerous lines of evidence (including detections of
extended X--ray emission, overdensities of galaxies in narrow--band images or
selected by infrared colour, cross--correlation analyses, and direct
spectroscopic studies of individual sources; see \cite{bes00} for a review)
indicate that at least some powerful radio sources are located at the centres
of galaxy overdensities. At higher redshifts, radio sources have been detected
out to $z=5.2$ \cite{bre99}, with some sources spectroscopically confirmed to
lie in overdense environments (cf. Kurk \etal\ \& Venemans \etal, in this
volume). This offers the prospect that AGN can be used to pinpoint galaxy
clusters back to the earliest epochs.

We have embarked upon a project to investigate in detail the environments of a
complete subsample of the most luminous radio sources from the equatorial BRL
sample of Best \etal\ \cite{bes99}, within the narrow redshift range $1.44 \le
z \le 1.7$. The fields of 6 sources have been imaged in the $R$, $J$ and $K$
wavebands, using the SOFI and SUSI-2 instruments on the NTT. This filter
combination was chosen to provide maximal sensitivity to old cluster
ellipticals with the $R$ and $J$ bands spanning the 4000\AA\ break and the $K$
waveband providing a long red baseline. The observations cover a field of view
of 5 by 5 arcmins ($\sim$2.7 by 2.7 Mpc at the radio source redshifts), and
reach limiting depths of $K \sim 20.6$, $J \sim 22.4$ and $R \sim 25.9$.
\vspace*{-0.1cm}

\section{Old red galaxies clustered around the AGN}
\vspace*{-0.1cm}

\begin{figure}
\begin{tabular}{cc}
\psfig{file=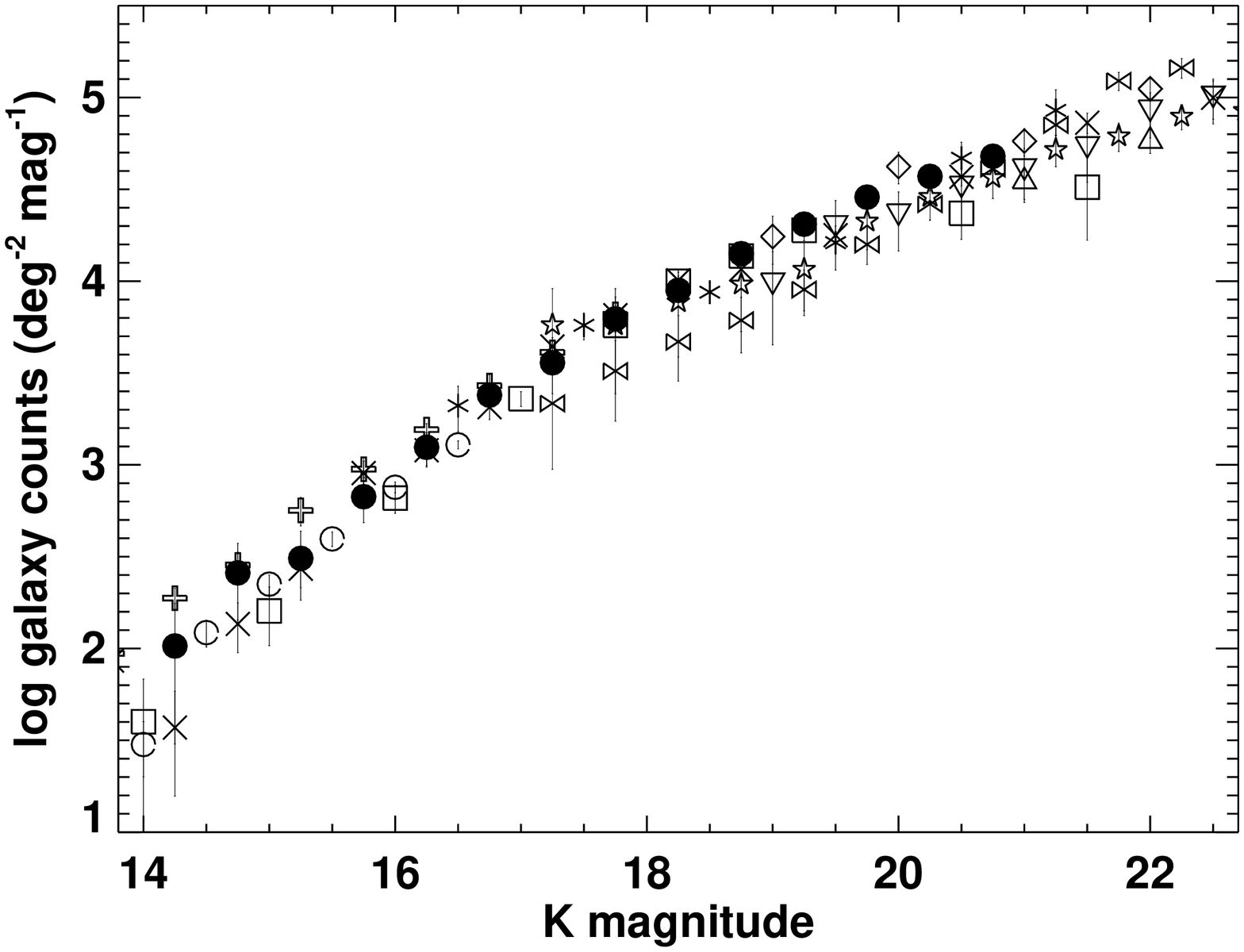,angle=90,width=6.7cm,clip=}
&
\psfig{file=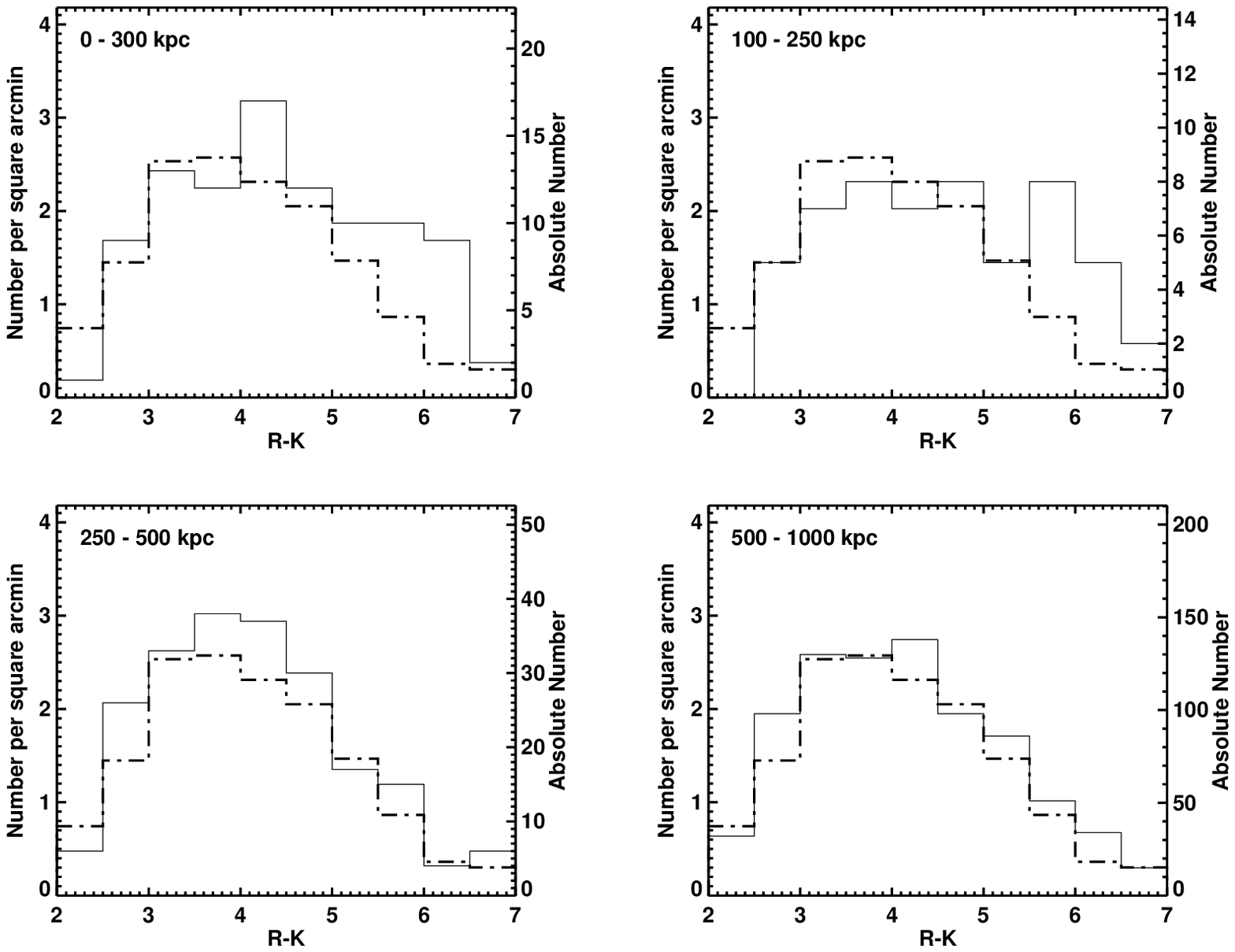,width=6.7cm,clip=}
\end{tabular}
\caption{\label{fig1} {\bf Left:} The K--band galaxy number counts per
magnitude per square degree (solid symbols) compared to various blank--field
surveys. {\bf Right:} The colour distribution of galaxies within 300\,kpc of
the AGN (solid line), compared with that at distances greater than 1\,Mpc
(dashed line).}
\end{figure}

The K--band number counts in the fields of these radio sources are compared
against the results of blank--field surveys in Figure~\ref{fig1} (left). The
derived counts are fully consistent with the literature counts at magnitudes
$K \lta 17.5$, which is the expected magnitude of brightest cluster galaxies
at $z \sim 1.5$. At fainter magnitudes the galaxy counts still lie within the
scatter of the other observations, but the counts in every magnitude bin from
$17.5 < K < 20.5$ exceed the literature average by at least 1$\sigma$. 

The $R$-$K$ colour distribution of the galaxies within 300\,kpc of the AGN is
displayed in Figure~\ref{fig1} (right), compared against the mean colour
distribution for galaxies at greater than 1\,Mpc radial distance from the AGN
in the same frames. There is a clear surplus of red ($R-K \gta 4$) galaxies.

\begin{figure}
\centerline{
\psfig{file=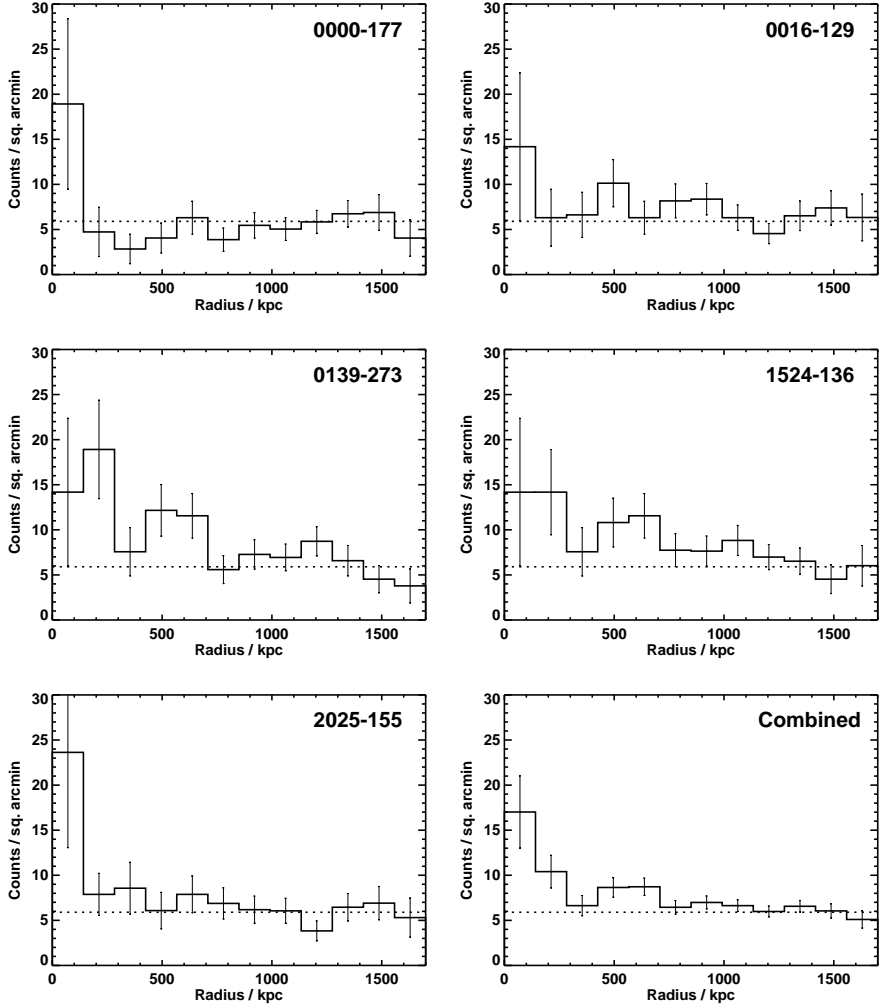,width=12cm,clip=} 
}
\caption{\label{fig2} The radial distribution of galaxies with $17.5 < K <
20.7$ and $R-K > 4$ in each of five fields separately, and the average of
these. The radius is calculated in kpc at the redshift of each AGN. The error
bars represent the Poissonian errors on the galaxy counts and the horizontal
dotted lines represent the blank--field expectation.}
\end{figure}

These results show that an excess of faint ($17.5 < K < 20.5$), red ($R-K \gta
4$) galaxies are found across the fields. These have the colours and
magnitudes expected of old passive galaxies at the radio source redshifts. The
radial distribution of the galaxies with these colours and magnitudes is
investigated in Figure~\ref{fig2}, for five of the radio source fields and for
a combined histogram produced by averaging across all five fields. Also
included on these plots is the expected number density for blank-field sources
of these magnitudes and colours; this was calculated from the K20 \cite{cim02}
and CADIS \cite{tho99} surveys. The results from combining the five AGN fields
show that red galaxy excesses are seen on two different scales around the
radio sources: there is a pronounced peak at small radii, $r \lta 150$\,kpc,
together with a larger scale overdensity extending across most of the
field. These results are in broad agreement with those of \cite{hal98};
however, with the larger radial coverage of the data provided here, it is now
possible to quantify the radial extent of the large--scale overdensity (which
extended beyond their field of view): the galaxy counts remain above field
expectations out to between 1 and 1.5\,Mpc radius.

Comparing the five individual fields in Figure~\ref{fig2}, however, it is
immediately apparent that this combined result is the average of five fields
with very different properties. 0016$-$129 and 2025$-$155 do both show radial
distributions similar to the average result, with pronounced peaks in the
inner bin followed by weak red galaxy overdensities to $\sim
1$\,Mpc. 0000$-$177 has a central pronounced peak of red galaxies, but shows
no large--scale excess (indeed quite the opposite). 0139$-$273 and 1524$-$136
are both much less centrally concentrated, with smoother central peaks and
much more gradual radial declines in red galaxy surface density out to the
blank--field expectations at radial distances of between 1 and 1.5\,Mpc. There
is clearly a large range in the nature, scale and significance of any red
galaxy overdensities around these AGN.

The Mpc--scale galaxy overdensities have the appearance of forming clusters of
galaxies. The wealth of galaxies around these AGN with the magnitudes and
colours expected of old elliptical galaxies at $z \sim 1.5$ indicate that the
epoch of elliptical galaxy formation must be earlier than this redshift; the
colours of these objects further suggest that the bulk of their stellar
populations must have formed at redshifts $z \gta 5$. The large size--scale of
these overdensities should not be surprising, since the timescale for objects
of density $\rho$ to separate from the Hubble Flow and collapse is given by $t
\sim (3 \pi / 32 G \rho)^{1/2}$ \cite{pea99}. For a proto-cluster, with $M
\sim 10^{14-15} M_{\odot}$ collapsing from 3-5\,Mpc radius, this corresponds
to a collapse time of several Gyr, and so such objects cannot yet be fully
collapsed by $z \sim 1.5$.

The small--scale overdensities have the size of compact groups of galaxies, or
of cluster cores. In the latter case, it is interesting that where
small--scale overdensities are seen, almost all of the galaxies that comprise
these have the red colours indicative of old passively evolving elliptical
galaxies. This indicates that morphological segregation may already have taken
place in the very inner regions around the AGN, implying that the
morphology--density relation [e.g. 22]\nocite{dre80} is imprinted into cluster
centres at a very early epoch.

In conclusion, substantial overdensities of red galaxies with the colours,
magnitudes and radial distribution expected of old passively evolving
elliptical galaxies have been found around radio sources at redshifts $z \sim
1.5$. Work is currently on--going to spectroscopically confirm the cluster
membership of these galaxies, and to age their stellar populations, which will
allow dramatic advances to be made in our understanding of how and when the
most massive elliptical galaxies form.

\end{document}